# Charge dynamics in doped triangular antiferromagnets

W. Q. Yu and Shiping Feng
*Department of Physics, Beijing Normal University, Beijing 100875, China*

Z. B. Huang and H. Q. Lin
*Department of Physics, The Chinese University of Hong Kong, Hong Kong, China*

Within the framework of the fermion-spin theory based on the charge-spin separation, the charge dynamics of the doped antiferromagnet on a triangular lattice is studied. The holon part is treated by using the loop expansion to the second order. It is shown that the charge dynamics is mainly caused by charged holons moving in the background of spinon excitations. The optical conductivity spectrum shows an unusual behavior at low energies and an anomalous midinfrared band, while the resistivity exhibits a nonlinearity in temperatures.

Over the last decade, much interest has focused on the nature of the two-dimensional (2D) doped antiferromagnets. This is motivated by the belief that the essential physics of the copper oxide materials is contained in the 2D doped Mott insulators.[1] After more than ten years of intense experimental studies of the copper oxide materials, it is clear that its normal state cannot be described by the conventional Fermi-liquid theory,[2] which perhaps is the most remarkable aspect of their unusual properties. The fascinating normal-state properties of the copper oxide superconductors, as manifested by the charge dynamics, result from special microscopic conditions:[1,2] (1) Cu ions situated in a square-planar arrangement and bridged by oxygen ions, (2) weak coupling between neighboring layers, and (3) doping in such a way that the Fermi level lies near the middle of the Cu-O $\sigma^*$ bond. One common feature of these compounds is the square-planar Cu arrangement. Howerer, it has been reported recently[3] that there is a class of the delafossite copper oxide materials, $RCuO_{2+\delta}$, with $R$ a rare-earth element, where the Cu ions site sit not on a square planar but on a triangular-planar lattice, therefore allowing a test of the geometry effect on the normal-state properties, while retaining some other unique microscopic features of the Cu-O bond. Other compounds with the triangular geometry have also been found,[4] such as experiments suggest the realization of the triangular spin-one-half antiferromagnet[5] in $NaTiO_2$. On the other hand, the doped antiferromagnet on the triangular lattice, where geometric frustration exists, is also of theoretical interest in their own right, with many unanswered fascinating theoretical questions.[6] Historically the undoped triangular antiferromagnet was first proposed to be a model for microscopic realization of the resonating-valence-bond (RVB) spin liquid due to the existence of the strong frustration.[7] This spin liquid state has been suggested by Anderson and Fazekas[7] to be the ground state of the undoped triangular antiferromagnet. Very soon after the discovery of the copper oxide superconductors, it has been argued[8] that the spin liquid state plays a crucial role in the superconductivity of the copper oxide materials. Since then several superconducting mechanisms based on the spin liquid state have been proposed. In particular, Kalmeyer and Laughlin[9] argued that the RVB state for the triangular lattice is apparently well described by a quantum Hall ground-state wave function for bosons, while the elementary excitations obey fractional statistics. Moreover, it has been shown[10] that the doped and undoped triangular antiferromagnets present a pairing instability in an unconventional channel. Therefore, it is very important to investigate the normal-state properties of this system by a systematic study.

Many researchers[8] have argued successfully that the $t$-$J$ model, acting on the Hilbert space with no doubly occupied site, provides a consistent description of the physical properties of the doped antiferromagnets. Within the $t$-$J$ model, the charge dynamics of the doped antiferromagnet on the square lattice in the underdoped and optimally doped regimes have been studied by a fermion-spin theory.[11] It was found that the optical conductivity shows non-Drude behavior at low energies and an anomalous midinfrared band in the charge-transfer gap, the resistivity exhibits a linear behavior in temperatures in the optimally doped regime and a nearly linear temperature dependence with deviations at low temperatures in the underdoped regime. It has been also shown that these unusual normal-state properties in the doped square antiferromagnet were caused by the strong electron correlation.[1,2,11] Since the strong electron correlation is common for both doped square and triangular antiferromagnets, it is expected that the unconventional charge dynamics observed on the square lattice may also be seen in the doped triangular antiferromagnet. In this paper, we study the charge dynamics of the doped triangular antiferromagnet within the $t$-$J$ model. We find that the midinfrared peak of the optical conductivity in the doped triangular antiferromagnets is severely depressed due to the geometry effect. As a consequence, the resistivity exhibits a nonlinear temperature dependence. We hope that the present study may induce further experimental work in the doped triangular antiferromagnet.

We start from the $t$-$J$ model defined on the triangular lattice which can be written as

$$H = -t\sum_{i\hat{\eta}\sigma} C_{i\sigma}^\dagger C_{i+\hat{\eta}\sigma} + \text{H.c.} - \mu\sum_{i\sigma} C_{i\sigma}^\dagger C_{i\sigma} + J\sum_{i\hat{\eta}} S_i \cdot S_{i+\hat{\eta}}, \qquad (1)$$

where the summation is over all sites $i$, and for each $i$, over its nearest neighbor $\hat{\eta}$, $C_{i\sigma}^\dagger$ ($C_{i\sigma}$) are the electron creation (annihilation) operators, $S_i = C_i^\dagger \sigma C_i/2$ are spin operators with $\sigma = (\sigma_x, \sigma_y, \sigma_z)$ as Pauli matrices, and $\mu$ is the chemical potential. The $t$-$J$ model (1) is subject to an important local constraint that double occupancy of a site by two electrons with opposite spins is not allowed. In spite of its simple form the $t$-$J$ model (1) proved to be very difficult to analyze, analytically as well as numerically, largely because of this electron single occupancy local constraint.[12] Recently a fermion-spin theory based on the charge-spin separation is proposed[13] to incorporate this local constraint. According to this theory we let $C_{i\uparrow} = h_i^\dagger S_i^-$ and $C_{i\downarrow} = h_i^\dagger S_i^+$, where the spinless fermion operator $h_i$ describes the charge (holon) degrees of freedom, while the pseudospin operator $S_i$ describes the spin (spinon) degrees of freedom. The main advantage of this approach is that the electron on-site local constraint for single occupancy can be treated exactly in analytical calculations. In this approach, the low-energy behavior of the $t$-$J$ model (1) in the fermion-spin representation can be rewritten as,[11]

$$H = -t \sum_{i\hat{\eta}} h_i h_{i+\hat{\eta}}^\dagger (S_i^+ S_{i+\hat{\eta}}^- + S_i^- S_{i+\hat{\eta}}^+) + \text{H.c.}$$
$$+ \mu \sum_i h_i^\dagger h_i + J_{eff} \sum_{i\hat{\eta}} S_i \cdot S_{i+\hat{\eta}}, \quad (2)$$

with $J_{eff} = [(1-\delta)^2 - \phi^2]$, where $\phi = \langle h_i^\dagger h_{i+\hat{\eta}} \rangle$ is the holon particle-hole order parameter. The Hamiltonian (2) contains the holon-spinon interaction, and therefore the spinon and holon are strongly coupled.

At the half-filling, the $t$-$J$ model is reduced to the Heisenberg model. In this case, the nature of the ground state, especially the existence of antiferromagnetic long-range order (AFLRO) for the antiferromagnetic (AF) Heisenberg model on the triangular lattice, has been extensively studied.[14–18] Recently, it has been shown[19] unambiguously that as in the square lattice, there is indeed AFLRO in the ground state of the AF Heisenberg model on the triangular lattice. However, it has been also shown[20] within the $t$-$J$ model that AFLRO in the doped square antiferromagnet vanishes around doping $\delta = 5\%$. Then AFLRO for the AF Heisenberg model on the triangular lattice should be decreased rapidly with increasing dopings than that on the square lattice due to the geometric frustration, so in both the underdoped and optimally doped regimes, there is no AFLRO for the doped triangular antiferromagnet, i.e., $\langle S_i^z \rangle = 0$. Therefore in the following discussions, we will study the motion of holons in the background of spin liquid state in the underdoped and optimally doped regimes. Not long ago, a self-consistent mean-field theory of the doped antiferromagnet on the square lattice in the underdoped and optimally doped regimes without AFLRO has been developed by one of us in Ref. 21. That approach should be applicable to the triangular lattice. Following discussions there, we can obtain the mean-field spinon and holon Green's functions $D^{(0)}(i-j, t-t') = \langle\langle S_i^+(t); S_j^-(t') \rangle\rangle_0$ and $g^{(0)}(i-j, t-t') = \langle\langle h_i(t); h_j^\dagger(t') \rangle\rangle_0$ of the doped triangular antiferromagnet as

$$D^{(0)}(k,\omega) = \frac{1}{2} B_k \left( \frac{1}{\omega - \omega(k)} - \frac{1}{\omega + \omega(k)} \right), \quad (3a)$$

$$g^{(0)}(k,\omega) = \frac{1}{\omega - \xi_k}, \quad (3b)$$

respectively, where $B_k = \Delta[(2\epsilon\chi_z + \chi)\gamma_k - (\epsilon\chi + 2\chi_z)]/\omega(k)$, $\Delta = 2ZJ_{eff}$, $\epsilon = 1 + 2t\phi/J_{eff}$, $\gamma_k = [\cos k_x + 2\cos(k_x/2)\cos(\sqrt{3}k_y/2)]/3$, $Z$ is the number of the nearest-neighbor sites, the mean-field holon excitation spectrum $\xi_k = 2\chi t Z \gamma_k - \mu$, and mean-field spinon excitation spectrum,

$$\omega^2(k) = \Delta^2 [\alpha(C_z - \epsilon\chi_z\gamma_k) - \alpha\epsilon\chi/(2Z) + (1-\alpha)/(4Z)]$$
$$\times (1 - \epsilon\gamma_k) + \Delta^2[\alpha\epsilon(C - \chi\gamma_k - 2\chi_z/Z)/2$$
$$+ \epsilon(1-\alpha)/(4Z)](\epsilon - \gamma_k),$$

with the spinon correlation functions

$$\chi = \langle S_i^+ S_{i+\hat{\eta}}^- \rangle, \quad C = (1/Z^2)\Sigma_{\hat{\eta}\hat{\eta}'} \langle S_{i+\hat{\eta}}^+ S_{i+\hat{\eta}'}^- \rangle,$$

$$\chi_z = \langle S_i^z S_{i+\hat{\eta}}^z \rangle, \quad C_z = (1/Z^2)\Sigma_{\hat{\eta}\hat{\eta}'} \langle S_{i+\hat{\eta}}^z S_{i+\hat{\eta}'}^z \rangle.$$

In order not to violate the sum rule $\langle S_i^+ S_i^- \rangle = 1/2$ when $\langle S_i^z \rangle = 0$, an important decoupling parameter $\alpha$ has been introduced, which is regarded as the vertex corrections.[21,22] The spinon correlation functions, holon order parameter, the decoupling parameter $\alpha$, and the chemical potential $\mu$ can be determined self-consistently. Within this mean-field theory, some physical properties of the doped triangular antiferromagnet have been calculated. At the half-filling, the spin liquid ground-state energy[23] per site is $E_g/NZJ = -0.966$, in very good agreement with those obtained by using the variational RVB state.[9,14] Away from the half-filling, the phase separation present in the doped square antiferromagnet,[24] is expected to be absent here. Moreover, the electron density of states (DOS) $\Omega(\omega)$ has been calculated,[25] and the result is shown in Fig. 1 with (a) the doping $\delta = 0.06$ and (b) $\delta = 0.12$ for the parameter $t/J = 2.5$ at zero temperature, where the existence of the normal-state gap in the electron DOS is an important feature. This result also shows that the electron DOS shifted towards higher energies with increasing doped holes, but the total weight is reduced. Since the integral of the electron DOS up to the Fermi energy has to be equal to the number of electrons, therefore there is a tendency that the normal-state gap narrows with the increase of doping since some states appear in the gap upon dopings. In comparison with the result of the doped square antiferromagnet,[21] it is shown that the normal-state gap in the doped triangular antiferromagnet is mainly induced by the spinon frustration.

At the mean-field level, the spinon and holon may be separated, but they are strongly coupled beyond the mean-field approximation due to many-body correlations. To study the charge dynamics, it is necessary to consider the fluctuations around the mean-field solution. In the framework of charge-spin separation, an electron is represented by the product of a holon and a spinon, then the external field can only be coupled to one of them. According to the Ioffe-Larkin combination rule,[26] the physical conductivity $\sigma(\omega)$ is given by

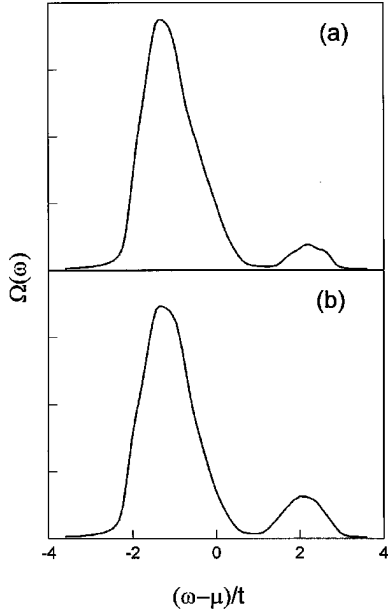

FIG. 1. Electron spectral density of the doped triangular antiferromagnet at the mean-field level in the doping (a) $\delta=0.06$ and (b) $\delta=0.12$ for the parameter $t/J=2.5$ in the zero temperature.

$$\sigma^{-1}(\omega)=\sigma_h^{-1}(\omega)+\sigma_s^{-1}(\omega), \quad (4)$$

where $\sigma_h(\omega)$ and $\sigma_s(\omega)$ are contributions from holons and spinons, respectively, and they can be expressed as,[27] $\sigma_h(\omega)=-\mathrm{Im}\,\Pi_h(\omega)/\omega$ and $\sigma_s(\omega)=-\mathrm{Im}\,\Pi_s(\omega)/\omega$, with $\Pi_h(t-t')=\langle\langle j_h(t);j_h(t')\rangle\rangle$ and $\Pi_s(t-t')=\langle\langle j_s(t);j_s(t')\rangle\rangle$ being the holon and spinon current-current correlation functions, respectively. In the present fermion-spin theory, the current densities of holons and spinons are obtained by taking the time derivatives of the polarization operator with the use of equations of motion, $j_h=2te\chi\Sigma_{i\hat{\eta}}\hat{\eta}h^\dagger_{i+\hat{\eta}}h_i$, $j_s=te\phi\Sigma_{i\hat{\eta}}\hat{\eta}(S_i^+S_{i+\hat{\eta}}^-+S_i^-S_{i+\hat{\eta}}^+)$, respectively. As in the case of a doped square antiferromagnet,[11] formally there is no direct contribution to the current-current correlation from spinons. However, because of the strong correlation between holons and spinons, the spinons' contribution is implicit through the spinon order parameters $\chi$, $\chi_z$, $C$, and $C_z$, which enter the holon part of contribution to the current-current correlation. The holon current-current correlation $\Pi_h(\omega)$ can be evaluated as

$$\Pi_h(i\omega_n)=-(2te\chi Z)^2\frac{1}{N}\sum_k\gamma_{sk}^2\frac{1}{\beta}\\\times\sum_{i\omega_n'}g(k,i\omega_n'+i\omega_n)g(k,i\omega_n'), \quad (5)$$

where $i\omega_n$ is the Matsubara frequency, $\gamma_{sk}^2=\{[\sin k_x+\sin(k_x/2)\cos(\sqrt{3}k_y/2)]^2+3[\sin(\sqrt{3}k_y/2)\cos(k_x/2)]^2\}/9$, and $g(k,i\omega_n)$ is the full holon Green's function. Previously, the second-order holon self-energy from the spinon pair bubble in the doped square antiferromagnet have been discussed in Ref. 11. Following their discussions, we obtain the second-order holon self-energy in the present doped triangular antiferromagnet as

$$\Sigma_h^{(2)}(k,\omega)=(\tfrac{1}{2}Zt)^2\frac{1}{N^2}\sum_{pp'}(\gamma_{p'-k}+\gamma_{p'+p+k})^2 B_{p'}B_{p+p'}\left(2\frac{n_F(\xi_{p+k})[n_B(\omega_{p'})-n_B(\omega_{p+p'})]-n_B(\omega_{p+p'})n_B(-\omega_{p'})}{\omega+\omega_{p+p'}-\omega_{p'}-\xi_{p+k}}\right.\\+\frac{n_F(\xi_{p+k})[n_B(\omega_{p+p'})-n_B(-\omega_{p'})]+n_B(\omega_{p'})n_B(\omega_{p+p'})}{\omega+\omega_{p'}+\omega_{p+p'}-\xi_{p+k}}\\\left.-\frac{n_F(\xi_{p+k})[n_B(\omega_{p+p'})-n_B(-\omega_{p'})]-n_B(-\omega_{p'})n_B(-\omega_{p+p'})}{\omega-\omega_{p+p'}-\omega_{p'}-\xi_{p+k}}\right), \quad (6)$$

where $n_F(\xi_k)$ is the Fermi-Dirac distribution function, while $n_B(\omega_k)$ is the Bose-Einstein distribution function. Then the full holon Green's functions are expressed as $g^{-1}(k,\omega)=g^{(0)-1}(k,\omega)-\Sigma_h^{(2)}(k,\omega)$. For the convenience of the discussion, this full holon Green's function can also be expressed as frequency integrals in terms of the spectral representation as

$$g(k,i\omega_n)=\int_{-\infty}^\infty\frac{d\omega}{2\pi}\frac{A_h(k,\omega)}{i\omega_n-\omega}, \quad (7)$$

with the holon spectral function $A_h(k,\omega)=-2\,\mathrm{Im}\,g(k,\omega)$. Substituting Eq. (7) into Eq. (5), and evaluating the frequency summations, we obtain the conductivity of the doped triangular antiferromagnet as

$$\sigma(\omega)=\frac{1}{2}(2te\chi Z)^2\frac{1}{N}\sum_k\gamma_{sk}^2\int_{-\infty}^\infty\frac{d\omega'}{2\pi}\\\times A_h(k,\omega'+\omega)A_h(k,\omega')\frac{n_F(\omega'+\omega)-n_F(\omega')}{\omega}. \quad (8)$$

In the underdoped and optimally doped regimes, the charge response, as manifested by the optical conductivity and resistivity, is a powerful probe for the systems of interacting electrons, and provides very detailed information of the excitations, which interact with carriers in the normal state. The optical conductivity (8) has been calculated numerically, and the zero temperature results with doping con-centration $\delta = 0.12$ (solid line), $\delta = 0.09$ (dashed line), and $\delta = 0.06$ (dotted line) for the parameter $t/J = 2.5$ are plotted in Fig. 2, where we take charge $e$ as the unit. Our results show that in the underdoped and optimally doped regimes, the characteristic features of the conductivity spectrum are (1) a low-energy peak at $\omega < 0.4t$, (2) considerable weights appear inside the charge-transfer gap of the undoped systems, defining the midinfrared band, and (3) the conductivity decays rapidly at low energies. Moreover, the gap in the conductivity is doping dependent, decreases with increasing dopings, and vanishes in the overdoped regime. In comparison with the results of the doped square antiferromagnet,[11] we find that although the qualitative shape of the optical conductivity spectrum of both doped square and triangular antiferromagnets seems to be similar, the midinfrared peak in the doped triangular antiferromagnet has been severely suppressed. Since the spectral weight from both the low-energy peak and midinfrared sideband represents the actual free-carrier density,[12] the suppression of the midinfrared band in the doped triangular antiferromagnet means that only a small amount of the free carrier is taken from the Drude absorption to the midinfrared band, which leads to the unusual decay of conductivity at low energies in the optimally doped regime. Unlike the doped square antiferromagnet, the optical conductivity for the doped triangular antiferromagnet cannot be fitted as $1/\omega$ as would be expected for the doped square antiferromagnet.

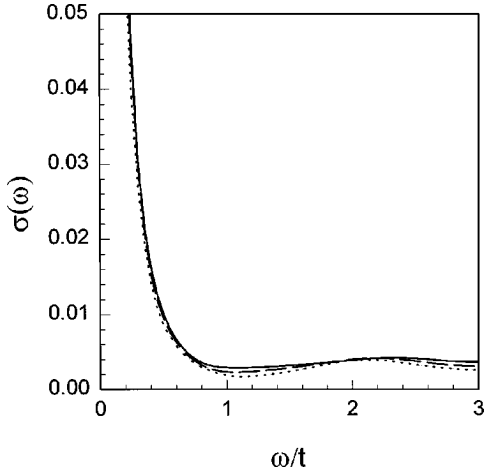

FIG. 2. The optical conductivity with doping concentration $\delta = 0.12$ (solid line), $\delta = 0.09$ (dashed line), and $\delta = 0.06$ (dotted line) at temperature $T = 0$ for the parameter $t/J = 2.5$.

The quantity which is closely related to the frequency-dependent conductivity (8) is the resistivity, which can be expressed as $\rho(T) = 1/\sigma_0(T)$, where the dc conductivity $\sigma_0(T)$ can be obtained from Eq. (8) as $\sigma_0(T) = \lim_{\omega \to 0} \sigma(\omega)$. This resistivity has been calculated numerically, and the results at doping concentration $\delta = 0.12$ and $\delta = 0.06$ for the parameter $t/J = 2.5$ are shown in Fig. 3(a) and Fig. 3(b), respectively. In contrast to the case of the doped square antiferromagnet,[11] the present results show that the resistivity of the doped triangular antiferromagnet exhibits a nonlinearity in temperatures.

Within the charge-spin separation framework, the charge dynamics is dominated by the scattering of charged holons, which are strongly renormalized because of the strong inter-actions with fluctuations of the surrounding spinon excitations. It has been shown that a remarkable property of the pseudogap in the doped antiferromagnet is that it appears in both spinon and holon excitations. For the doped square antiferromagnet, the linear behavior in temperatures in the optimally doped regime and a nearly linear temperature dependence with deviations at low temperatures in the underdoped regime in the resistivity are connected with the holon pseudogap behavior,[28] since the holon pseudogap in the doped square antiferromagnet is doping and temperature dependent, and grows monotonically as the doping $\delta$ decreases and disappears in the optimally doped regime. Moreover, this holon pseudogap also decreases with increasing temperatures, and vanishes at higher temperatures.[28] But in contrast to the case of the doped square antiferromagnet, the holon pseudogap in the doped triangular antiferromagnet is almost doping and temperature independent, which can be understood from the physical property of the holon DOS $\Omega_h(\omega) = 1/N \Sigma_k A_h(k,\omega)$. We have performed a numerical calculation for the holon DOS, and the result at the doping $\delta = 0.06$ and $\delta = 0.12$ for the parameter $t/J = 2.5$ at the temperature $T = 0$ are shown in Figs. 4(a) and 4(b), respectively. For comparison, the corresponding mean-field result (dashed line) is also shown in Fig. 4. The holon DOS in the underdoping $\delta = 0.06$ as a function of energy for the temperature (a) $T = 0.1J$ and (b) $T = 0.5J$ are plotted in Fig. 5. These results show that the holon DOS in the mean-field approximation (MFA) contain the central part only, which comes from the noninteracting particles. After including fluctuations the central peak is renormalized as two peaks separated by the holon pseudogap. This holon pseudogap in the doped triangular antiferromagnet persists in the overdoped regime even at higher temperatures, and is closely related to the strong spinon frustration. This holon pseudogap would also reduce the holon scattering and thus is responsible for the nonlinear dependence in temperatures of the resistivity in the doped triangular antiferromagnet. In particular, we also note that the holon and spinon dispersions in the 2D $t$-$J$ model on the triangular lattice are very similar to those in the 2D $t$-$t'$-$J$ model on the square lattice, whereas the latter has been argued[29] as the proper model of the copper oxide materials in the underdoped regime. On the other hand, the geometric effect (then spin frustration effect) is induced by the structure distortion of the $CuO_2$ planes in the orthorhombic structure in the underdoped regime,[30] and some anomalous properties are closely related to this lattice distortion. This is also why the behaviors of the present optical conductivity and resistivity in the doped triangular antiferromagnet are qualitatively consistent with the copper oxide materials in the underdoped regime.

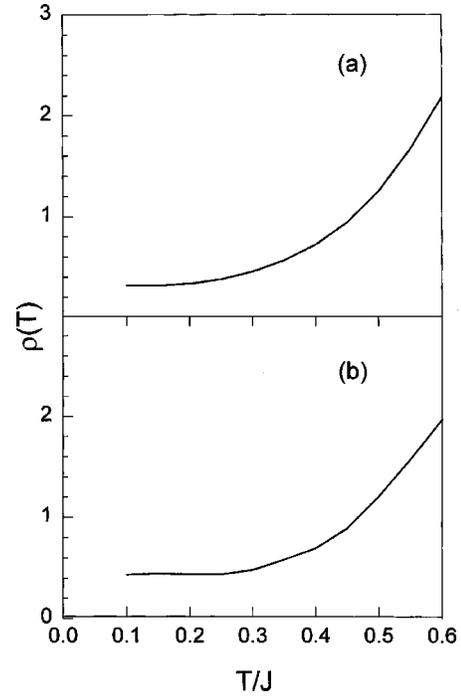

FIG. 3. The resistivity in the doping concentration (a) $\delta = 0.12$ and (b) $\delta = 0.06$ for the parameter $t/J = 2.5$.

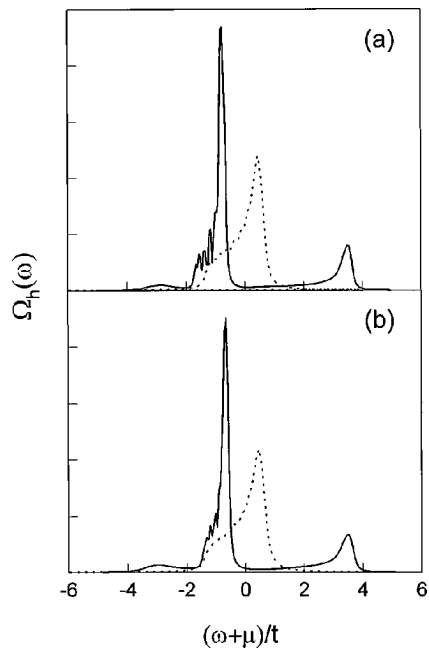

FIG. 4. The holon density of states at $t/J=2.5$ for (a) the doping $\delta=0.06$ and (b) $\delta=0.12$ in the zero temperature. The dashed line is the result at the mean-field level.

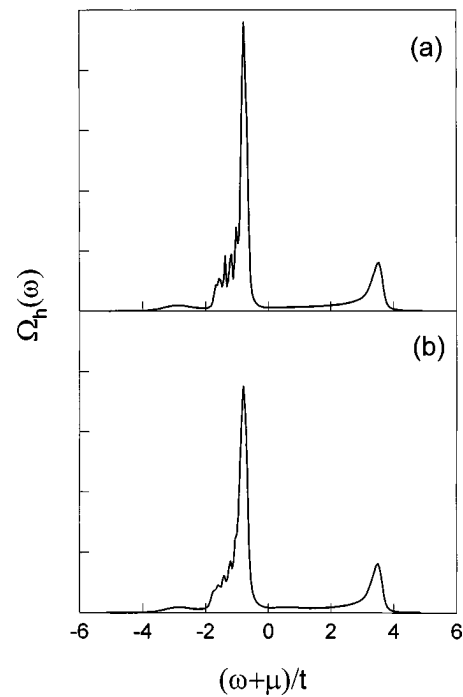

FIG. 5. The holon density of states for $t/J=2.5$ at the doping $\delta=0.06$ in the temperature (a) $T=0.1J$ and (b) $T=0.5J$.

In summary, we have discussed the charge dynamics of the doped triangular antiferromagnet in the underdoped and optimally doped regimes within the framework of the fermion-spin theory based on the charge-spin separation. The holon part has been treated by using the loop expansion to the second order. Our results show that the charge dynamics is mainly caused by charged holons moving in the background of spinon excitations. The optical conductivity spectrum shows an unusual behavior at low energies and an anomalous midinfrared band, while the resistivity exhibits a nonlinearity in temperatures.

The authors would like to thank Professor C. D. Gong and Professor J. Li for stimulating discussions and Feng Yuan and Xianglin Ke for their help in the numerical calculations. This work was supported by the National Natural Science Foundation under Grant No. 19774014 and the Earmarked Grant for Research from the Research Grants Council (RGC) of Hong Kong SAR.